\documentclass[aps,prl,twocolumn,superscriptaddress,showpacs]{revtex4}

\usepackage[dvips]{graphicx}
\usepackage{latexsym}

\begin{document}

\title{
Field dependence of the zero energy density of states \\
around vortices in an anisotropic gap superconductor
}

\author{
N. Nakai
}
\email[]{nakai@yukawa.kyoto-u.ac.jp}
\affiliation{
Yukawa Institute for Theoretical Physics, Kyoto University, Kyoto 606-8502, Japan
}
\author{
P. Miranovi\'{c}
}
\affiliation{
Department of Physics, University of Montenegro, Podgorica 81000, Serbia and Montenegro
}
\author{
M. Ichioka
}
\affiliation{
Department of Physics, Okayama University, Okayama 700-8530, Japan
}
\author{
K. Machida
}
\affiliation{
Department of Physics, Okayama University, Okayama 700-8530, Japan
}

\date{\today}

\begin{abstract}
It is shown theoretically that the Sommerfeld coefficient 
$\gamma(B)$ in the mixed state contains useful information on the gap
topology in a superconductor. We accurately evaluate $\gamma(B)$
against magnetic induction $B$ by using microscopic quasi-classical theory.
The linear portion in \mbox{$\gamma(\propto B)$}
relative to $B_{\rm c2}$ 
and its slope in low field faithfully corresponds to the gap anisotropy, 
so that we can distinguish the nodal gap and the 
anisotropic full gap, and estimate the magnitude 
of the gap anisotropy. 
\end{abstract}


\maketitle

There has been much attention focused 
on exotic superconductors in recent years, 
because various new types of
superconductors are successfully synthesized.
For example, high T$_{\rm c}$ cuprates and CeCoIn$_5$ are thought 
that the superconducting state with $d$-wave pairing is realized. 
In the case of UPt$_3$ \cite{machida}
and Sr$_2$RuO$_4$ \cite{mackenzie}, 
the superconducting state is thought 
to be a spin triplet pairing state. 
Only for a few superconductors, however, the realized pairing function
has been conclusively determined as for the gap structure 
and its parity.
Needless to say, it is crucial to know these when we understand
its pairing mechanism.

One of the standard methods to resolve the gap structure is 
to use the temperature ($T$) dependence of physical quantities, 
such as the specific heat $C(T)$, 
the thermal conductivity $\kappa(T)$ and the spin lattice relaxation time $T_1^{-1}(T)$.
These comprise a unique set of characteristic $T$-dependences, 
allowing us to extract the gap topology; In the case of the nodal gap,
where the gap in momentum space is zero along a line on the Fermi surface,
\mbox{$C(T)\propto T^2$},
\mbox{$\kappa(T)\propto T^2$}
and \mbox{$T_1^{-1}(T)\propto T^3$} \cite{sigrist}.
In contrast, these are all activation type for full gap case.
It is often the case that these $T$-dependences are masked by 
other contributions such as impurities, phonon, nuclear spin, etc. 
Thus it is difficult to definitely 
conclude the gap anisotropy by the $T$-dependence of these quantities alone.
We certainly need another methods to supplement this situation on top of existing ones.

The magnetic field ($B$) in a superconductor adds a new dimension:
The specific heat and thermal conductivity measurements 
in the mixed state
as a function of $B$ give rise to 
additional information on low-lying excitation 
spectrum induced around vortices, 
which sensitively reflects the gap structure \cite{ichioka1,ichioka2}.
In fact it has been already demonstrated that the orientational 
Sommerfeld coefficient $\gamma(\theta)$ 
and thermal conductivity $\kappa(\theta)$ show 
a characteristic oscillation with the relative angle $\theta$ 
between the field direction 
and the direction
of the gap node (or minimum) 
\cite{vehkter,maki,pedja,izawa,tuson,aoki,deguchi}. 
This allows us to precisely determine the gap anisotropy.

As for observation of low-lying quasi-particle excitations in the mixed state,
scanning tunneling microscopy (STM) provides a direct way 
to probe the spatial extension 
of these quasi-particles \cite{sakata}. 
STM experiments on \mbox{2$H$-NbSe$_2$} 
comprise rich information on the gap structure of this system~\cite{hess}. 
However in this method, as well as 
angle-resolved photoemission spectroscopy (ARPES) 
which yields momentum space information~\cite{damascelli,yokoya,kiss},
observable materials are restricted due to the surface-sensitivity. 

Thus it is quite desirable to establish another method to  probe the gap structure through bulk measurements. 
Here we propose the field-dependence of specific heat measurement  at low temperature, 
namely the Sommerfeld coefficient $\gamma(B)$ in the mixed state.
We demonstrate that the precise $\gamma(B)$ measurement 
at low field yields an important piece of
information on the degree of the gap anisotropy, 
or the ratio of the minimum gap
and maximum gap in a superconductor.

This type of experiments has been done in the past 
on a variety materials
such as Nb \cite{ferreira}, 
\mbox{2$H$-NbSe$_2$} \cite{hanaguri,boaknin2},
\mbox{CeRu$_2$} \cite{hedo}, 
\mbox{LuNi$_2$B$_2$C} \cite{nohara},
\mbox{YBa$_2$Cu$_3$O$_7$} \cite{wang},
\mbox{Sr$_2$RuO$_4$} \cite{deguchi}
and \mbox{Pr$_{2-x}$Ce$_x$CuO$_{4-\delta}$} \cite{balci}. 
However, because of lack of theoretical detailed calculations, 
these data remain largely
to be analyzed.
Theoretically $\gamma(B)$  was interpreted 
in terms of ``rigid normal core concept'' \cite{deGennestext},
that is, \mbox{$\gamma(B)\propto B/B_{\rm c2}$} 
for superconductors with full gap 
because each flux carries 
a certain zero-energy density of states (ZDOS).
Volovik succeeded in describing 
$qualitatively$ $\gamma(B)$ (\mbox{$\propto \sqrt B$})
for the nodal gap superconductor \cite{volovik}. 
However, there are many experimental data that  $\gamma(B)$ shows 
convex curve, including $\sqrt{B}$-like behavior, 
even in full-gap superconductors. 
So far no one has studied this problem in a $quantitative$
level which allows us 
to examine abundant experimental data $\gamma(B)$ 
in order to deduce the anisotropy of the gap structure, 
or the ratio of the minimum and 
maximum gaps, including the nodal case 
where the minimum gap is zero.

The quasi-classical Eilenberger framework \cite{eilenberger}
is most suited for the present purpose,
which is microscopic valid 
for almost all superconductors where $k_F\xi\gg 1$.
We assume a superconductor 
in the clean limit and with the cylindrical Fermi surface,
which does not alter our main conclusion 
essentially valid for three-dimensional
case too.

We introduce the pair potential
\mbox{$\Delta({\bf r}, \theta)$}, 
the vector potential 
\mbox{${\bf A}({\bf r})$}
and the quasi-classical Green's functions
\mbox{$g({\rm i}\omega_n, {\bf r}, \theta)$}, 
\mbox{$f({\rm i}\omega_n, {\bf r}, \theta,)$}
and \mbox{$f^{\dagger}({\rm i}\omega_n, {\bf r}, \theta)$},
where ${\bf r}$ is the center of mass coordinate of the Cooper pair.
The direction of the relative momentum of the Cooper pair,
\mbox{$\hat{\bf k}={\bf k}/\left|{\bf k}\right|$};
\mbox{$\hat{\bf k}=(\cos\theta, \sin\theta)$},
is denoted by the polar angle $\theta$
relative to ${\bf x}$-direction.
The Eilenberger equation is given by
\begin{eqnarray}
\left\{\omega_n+\frac{\rm i}{2}{\bf v}_{\rm F}\cdot\left(\frac{\nabla}{\rm i}
+\frac{2\pi}{\phi_0}{\bf A}({\bf r})
\right)\right\}f({\rm i}\omega_n, {\bf r}, \theta)\nonumber\\
=\Delta({\bf r}, \theta)g({\rm i}\omega_n, {\bf r}, \theta),
\label{eq1} 
\\ 
\left\{\omega_n-\frac{\rm i}{2}{\bf v}_{\rm F}\cdot\left(\frac{\nabla}{\rm i}
-\frac{2\pi}{\phi_0}{\bf A}({\bf r})
\right)\right\}f^{\dagger}({\rm i}\omega_n, {\bf r}, \theta)\nonumber\\
=\Delta^{\ast}({\bf r}, \theta)g({\rm i}\omega_n, {\bf r}, \theta),
\\ 
g({\rm i}\omega_n, {\bf r}, \theta)
=[1-f^{\dagger}({\rm i}\omega_n, {\bf r}, \theta)
f({\rm i}\omega_n, {\bf r}, \theta)]^{\frac{1}{2}}, 
\label{eq3} 
\end{eqnarray}
where
\mbox{${\rm Re}\, g({\rm i}\omega_n, {\bf r}, \theta)>0$}
and
\mbox{${\bf v}_{\rm F}=v_{\rm F}\hat{\bf k}$}
is the Fermi velocity.
$\phi_0$ is flux quantum.

The external field ${\bf H}$ is applied along ${\bf z}$-direction.
With the symmetric gauge,
the vector potential is written as
\mbox{${\bf A}({\bf r})=
(1/2){\bf H}\times{\bf r}+{\bf a}({\bf r})$}.
Here ${\bf a}({\bf r})$ is related to
the internal field ${\bf h}({\bf r})$,
where \mbox{${\bf h}({\bf r})=\nabla\times{\bf a}({\bf r})$}.
The pairing interaction is assumed separable
\mbox{$V(\theta, \theta')=V_0\phi(\theta)\phi(\theta')$}
so that the pair potential is
\mbox{$\Delta({\bf r},\theta)=\Delta({\bf r})\phi(\theta)$}.

We numerically~\cite{ichioka1,ichioka2,pedja} solve Eqs. (\ref{eq1})-(\ref{eq3}) with 
the self-consistent conditions for
\mbox{$\Delta({\bf r},\theta)$} 
and
\mbox{${\bf a}({\bf r})$}; 
\begin{eqnarray}
\Delta({\bf r})
=N_0V_02\pi T\sum^{\omega_c}_{\omega_n>0}\int_0^{2\pi}\frac{d\theta '}{2\pi}
\phi(\theta')f({\rm i}\omega_n, {\bf r}, \theta '),\\
{\bf j}({\bf r})=-\frac{\pi\phi_0}{\kappa^2\Delta_0\xi^3}
2\pi T\sum^{\omega_c}_{\omega_n>0}\int_0^{2\pi}\frac{d\theta}{2\pi}
\frac{\hat{\bf k}}{\rm i}g({\rm i}\omega_n, {\bf r}, \theta),
\end{eqnarray}
where 
${\bf j}({\bf r})=\nabla\times\nabla\times {\bf a}({\bf r})$, 
$N_0$ is the density of states
at the Fermi energy in the normal state.
The cut-off energy is set as \mbox{$\omega_{\rm c}=20T_{\rm c}$}.
\mbox{$\kappa=\sqrt{7\zeta (3)/72}
(\Delta_0/T_{\rm c})\kappa_{\rm GL}$}.
$\zeta (3)$ is Riemann's zeta function.
The Ginzburg-Landau parameter is set as \mbox{$\kappa_{\rm GL}=9.0$}.
$\Delta_0$ is the uniform gap at $T=0$.

The local DOS of the energy $E$ is given by
\begin{eqnarray}
N(E, {\bf r})=N_0\int_0^{2\pi}\frac{d\theta}{2\pi }
{\rm Re}\, g({\rm i}\omega_n\to E+{\rm i}\eta,{\bf r}, \theta),
\end{eqnarray}
where $g$ is calculated by the above Eilenberger equations
with \mbox{${\rm i}\omega_n\to E+{\rm i}\eta$},
using the solution $\Delta$ and $A({\bf r})$
by the self-consistent calculation.
We typically use \mbox{$\eta=0.01\Delta_0$}.

The calculation for Green's functions is performed 
at $T/T_{\rm c}=0.1$
within the vortex lattice unit cell,
which is divided into \mbox{$81 \times 81$} mesh points.
In this study, we assume that vortices form a triangular lattice.
One of the nearest neighbor vortices is situated along ${\bf x}$-direction.

The gap function is taken as \mbox{$\phi(\theta)=1$} 
for the isotropic gap case,
and 
\mbox{$\left|\phi(\theta)\right|=\sqrt{1-\cos 6\theta}=\sqrt{2}\left|\cos 3\theta \right|$}
for the line-node gap case.
The gap function with finite minimum gap 
is chosen as
\mbox{$\phi(\theta)=(1-\alpha \cos 6\theta)/\sqrt{1+\alpha^2/2}$},
where $\alpha$ denotes the degree of the gap anisotropy,
whose sign and magnitude characterize the gap structure.
Here hexagonal crystal symmetry is assumed.

The total ZDOS  or $N(B)$ is the spatial average 
of \mbox{$N(E=0, {\bf r})$},
which is given by
\mbox{$N(B)=\int N(E=0, {\bf r})d{\bf r}/\int d{\bf r}$}.
This is related to $\gamma(B)$ at $T\rightarrow 0$ limit.
To characterize delocalized contribution  to total ZDOS,
we define ``extended ZDOS'' by
\mbox{$N_{\rm ext}(B)
=\int_c N(E=0, {\bf l})d{\bf l}/\int_c d{\bf l}$},
where the path of the integration 
\mbox{$\int_c \cdots d{\bf l}$}
is along the boundary of the Wigner-Seitz cell. 
Thus $N_{\rm ext}(B)$ is a useful quantity, 
related to the states outside of the vortex core. 
This is directly measurable by STM,
and would somehow be related to the thermal conductivity.

\begin{figure}[tbp]
\includegraphics{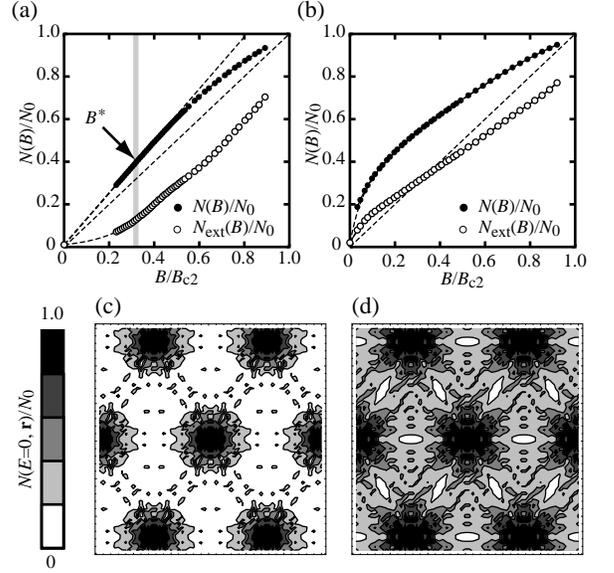}%
\caption{
\label{fig1}
(a) and (b) show field dependences of ZDOS. 
The filled (open) circles are for the total (extend) ZDOS.
In (c) and (d) density maps of ZDOS are shown around
a vortex core at the center and neighbor vortex cores. 
$B/B_{\rm c2}=0.24$.
The density height of the vortex center is truncated by 1.
Isotropic gap case ((a) and (c)) and line-node gap case ((b) and (d)) 
are presented. 
}
\end{figure}
\begin{figure*}[tb]
\includegraphics{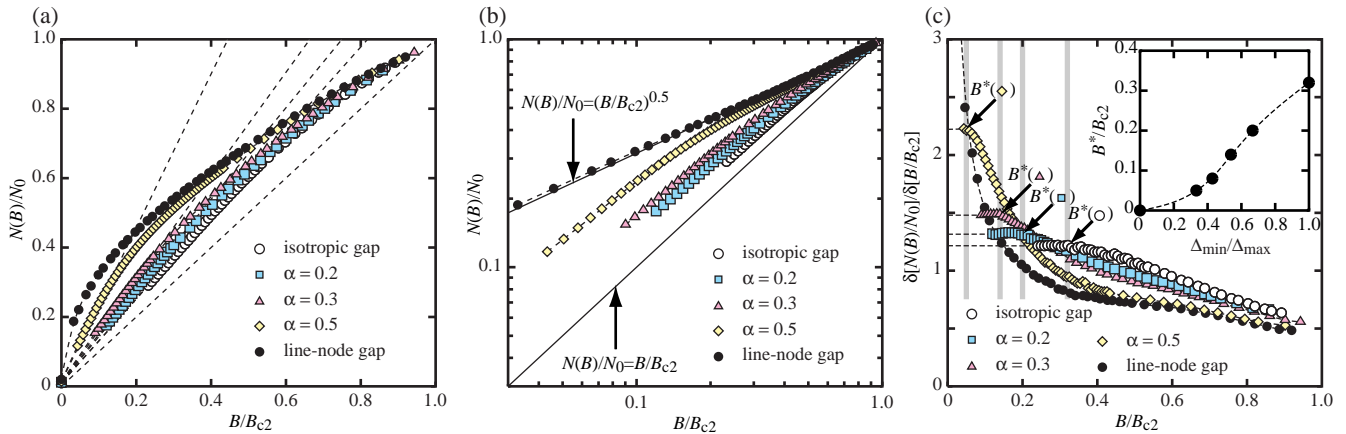}%
\caption{
\label{fig2}
(color)
(a) Field dependence of $N(B)$ for anisotropic gap cases.
Plotted data are for 
\mbox{$\alpha=0, 0.2, 0.3, 0.5$}
and the line-node 
from bottom to top. 
(b) The same data are replotted in
logarithmic scale.
(c) Field dependence of derivative $\delta N(B)/ \delta B$,  
i.e., slope of $N(B)$. 
From right to left 
shadow lines indicate each $B^\ast$
for $\alpha=0, 0.2, 0.3, 0.5$.
In the inset, $B^{\ast}/B_{\rm c2}$ is plotted
as a function of  $\Delta_{\rm min}/\Delta_{\rm max}$. 
$\Delta_{\rm min}=\phi(\theta=0)$ 
($\Delta_{\rm max}=\phi(\theta=\pi/6)$) is 
the minimum (maximum)
of the gap function $\phi(\theta)$. 
}
\end{figure*}
Let us now show our results deduced
from extensive self-consistent calculations. 
We first take up two extreme cases for the isotropic gap
and line node gap:
In \mbox{Fig. 1 (a)}, the field dependence of the ZDOS
for the isotropic gap is shown. 
It is seen that $N(B)$
is linear for smaller induction field $B$
up to a certain field $B^\ast$.
Beyond $B^\ast$ $N(B)$ becomes curving
up to the upper critical field $B_{\rm c2}$.
As for the extended ZDOS $N_{\rm ext}(B)$,
it is concaved all the way up to $B_{\rm c2}$.

The presence of $B^\ast$ can be understood as follows:
As $B$ increases, considerable overlapping of ZDOS 
at each core makes the linear behavior in $N(B)$ change 
to non-linear at $B^\ast$. 
Up to $B^\ast$, vortices carry independently
a certain ZDOS to add up. We see it from \mbox{Fig. 1 (c)}
that ZDOS is quite small in a large part of unit cell
except for the small core region.
This is compared with that for line node case in \mbox{Fig. 1 (d)}. 
It is emphasized that the low field
linear behavior in $N(B)$ superficially accords with the so-called
rigid normal core model \cite{deGennestext}, 
but the situation is more interesting and subtle:
(1) There exists  a crossover field $B^\ast$ from linear to non-linear
even in the isotropic gap case, 
which is a key fact important in the anisotropic case below. 
(2) Even in small $B$ region $N_{\rm ext}(B)$ is substantial,
meaning that vortex cores are overlapped there.
This invalidates rigid normal core picture \cite{deGennestext}.

The line node gap case as an extreme anisotropic one
is shown in \mbox{Figs. 1 (b) and (d)}.
Both curves of $N(B)$ and $N_{\rm ext}(B)$ are convex
with no linear part at low $B$.
These results are totally different
from those of the isotropic case.
In particular, the small field behavior is affected 
by the presence of nodes.
In the nodal gap case
$N_{\rm ext}(B)$ increases rapidly with $B$ as seen from \mbox{Fig. 1 (b)}, 
because the quasi-particle wave functions with zero-energy extend 
to far outside of the vortex core due to the node (see \mbox{Fig. 1 (d)}). 
Therefore $N_{\rm ext}(B)$ shows the convex behavior at lowest $B$, 
which is contrasted with the concave behavior of $N_{\rm ext}(B)$ 
at small $B$ in \mbox{Fig. 1 (a)}. 
Moreover $N(B)$ is non-linear from the lowest field
because the quasiparticles near the gap node are free 
from confinement in the core region, as 
clearly seen in \mbox{Figs. 1 (b) and (d)}.

These results again seem to coincide superficially 
with Volovik's Doppler shift calculation \cite{volovik}.
Here we emphasize that Volovik calculation
neglects the core contribution coming from the
localized quasi-particles, taking into account
the extended quasi-particle contribution.
Volovik result ($N(B)\propto \sqrt B$) is, strictly speaking,
only valid for smallest $B$ region and is qualitative
in nature.

We now come to our main results for general anisotropic gap case;
In \mbox{Figs. 2 (a) and (b)}, 
we display $N(B)$ for various values of $\alpha$,
including the above extreme cases with line node.
It is seen from these that 
(A) For $\alpha\neq 1$ $N(B)$ is linear in $B$ at lower fields.
(B) The linear $B$ region becomes limited to lower $B$ 
as $\alpha$ increases.
This can be easily seen in logarithmic plot of \mbox{Fig. 2 (b)},
where the slopes (\mbox{$\alpha \ne 1$}) in logarithmic plot change
from \mbox{$N(B)/N_0\propto B/B_{\rm c2}$}
to \mbox{$(B/B_{\rm c2})^{0.5}$} as $B$ increases.
There exists always a crossover field $B^{\ast}$ for
general $\alpha$, whose precise value is determined below.
(C) At higher field near $B_{\rm c2}$, 
$N(B)$ behaves similarly, 
independent of $\alpha$ values.
It is clear that $N(B)$ is not described 
by a single exponential 
form \mbox{$N(B) \propto B^{p}$}
covering all $B$ region except for the nodal gap case. 
Near $B_{\rm c2}$ the exponent $p$ is approximated 
to be 0.5 for all curves.

These results are interpreted along the same line above,
namely, in terms of quasi-particle's localized-delocalized
crossover upon varying $B$.
In the region $B<B^{\ast}$ each vortex core independently 
contributes to the DOS at zero energy while $B>B^{\ast}$
the quasi-particles are brought with flux lines overlapping each other,
making $N(B)$ non-linear. Thus a question is how $B^{\ast}$
is correlated to the gap anisotropy $\alpha$.

We exhibit the derivative $\delta N(B)/\delta B$ in \mbox{Fig. 2 (c)}. 
If $N(B)$ is linear in $B$, the derivative is a constant and 
equals to the slope of $N(B)$ near $B=0$. 
This linear slope constant, 
indicating how much DOS is brought into a system by a flux line, 
increases with $\alpha$. 
This implies that the 
ZDOS per vortex increases with the gap anisotropy.
Physically it is because the average core radius measured by
the extension of ZDOS around a core is effectively larger for larger 
gap anisotropy. 

It is clearly seen from \mbox{Fig. 2 (c)}
that there is a constant $\delta N(B)/\delta B$ region 
up tp $B^\ast$ for each value of $\alpha$. 
As shown in the inset of \mbox{Fig. 2 (c)},
this crossover field $B^{\ast}$ progressively 
decreases as the ratio of the minimum and maximum gaps  grows,
or $\alpha$ increases towards 1. 
In the nodal gap case, 
the derivative  $\delta N(B)/ \delta B$ is diverging 
because the exponent $p$ in $N(B)$ is less than unity \cite{node}. 
Thus it is possible 
to estimate the degree of the gap anisotropy $\alpha$ by 
identifying $B^{\ast}$ through specific heat measurement.

\begin{figure}[tbp]
\includegraphics{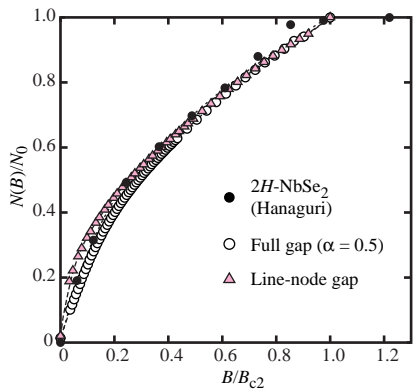}%
\caption{
\label{fig3}
(color)
Comparison with the $\gamma(B)$ in 2$H$-NbSe$_2$ \cite{hanaguri}
with our results for $\alpha=0.5$ and line node gap.
}
\end{figure}
For illustrative purpose, we show the comparison between
the $\gamma(B)$ data of \mbox{2$H$-NbSe$_2$} \cite{hanaguri}
and our results (\mbox{$\alpha=0.5$} and line node case) 
in \mbox{Fig. 3}. 
\mbox{2$H$-NbSe$_2$} is a anisotropic but full gap superconductor. 
While it is possible even to say that $\gamma(B) \propto \sqrt{B}$ 
at high fields, $\gamma(B) $ deviates from $\sqrt{B}$-behavior 
at low fields. 
The low field data fit to 
\mbox{$\alpha=0.5$} rather than line node gap. 
The anisotropy \mbox{$\alpha=0.5$} 
is also estimated by the recent ARPES data \cite{kiss} and 
also by our previous identification \cite{hayashi}.

In order to identify the $\alpha$ value in the actual measurement,
we must examine the applicability of the present approach.
The possible obstacles, which are not considered here are following:
(1) Non-magnetic impurities certainly modify $N(B)$ behavior
as discussed elsewhere \cite{pedja2,kita}. 
One needs samples in clean limit.
(2) The actual superconductors are often multi-band system,
or multi-gap one. Typical example is MgB$_{2}$ where distinct
two different gaps are assigned for two-bands. 
In fact $N(B)$ consists of two linear lines \cite{bouquet}, 
each segment is explained in terms of isotropic gap \cite{nakai, ichioka3}.
Thus it is possible to separate out the multi-gap effect.
Such multi-gap behavior is seen in various systems,
e.g. \mbox{Sr$_{2}$RuO$_{4}$} \cite{deguchi},
\mbox{2$H$-NbSe$_2$} \cite{boaknin2}
or \mbox{CeRu$_2$} \cite{hedo} etc.
For \mbox{2$H$-NbSe$_2$}, other measurements \cite{yokoya,kiss}
imply multi gaps.
(3) Since there exists the Fermi velocity anisotropy,
one may concern its effect on the present conclusion.
It turns out 
that it does not alter our conclusion essentially \cite{nakai2}.
(4) Although we considered here 2D system, 
3D effect also is checked. 
We find that it does not also change our conclusion
because $N(B)$ is a bulk quantity.
Thus we believe 
that a careful $N(B)$ measurement at low enough $T$ 
is a powerful method to identify the gap anisotropy.

\begin{acknowledgments}
We are grateful to Y. Matsuda, T. Sakakibara, K. Izawa, T. Hanaguri,
T. Kiss and J. E. Sonier
for useful discussions.
One of the authors (N. N.) is supported 
by a Grant-in-Aid for the 21st Century COE 
``Center for Diversity and University in Physics".
\end{acknowledgments}
%
\bibliography{basename of .bib file}

\begin{thebibliography}{99}
\bibitem{machida}
 K. Machida  et al., 
 J. Phys. Soc. Jpn., {\bf 58}, 4116 (1989);
ibid, {\bf 68}, 3364 (1999).
\bibitem{mackenzie}
 For a review,
 see A. P. Mackenzie and Y. Maeno
 Rev. Mod. Phys. {\bf 75}, 657 (2003).
\bibitem{sigrist}
 For example,
 see M. Sigrist and K. Ueda, Rev. Mod. Phys.
 {\bf 63}, 239 (1991).
 \bibitem{ichioka1}
 M. Ichioka {\it et al}.,
 Phys. Rev. B {\bf 55}, 6565 (1997).
\bibitem{ichioka2}
 M. Ichioka {\it et al}.,
 Phys. Rev. B {\bf 59}, 184 (1999);
 ibid, {\bf 59}, 8902 (1999).
\bibitem{vehkter}
 I. Vekhter {\it et al}.,
 Phys. Rev. B {\bf 59} 9023 (1999).
\bibitem{maki}
 K. Maki {\it et al},
 Phys. Rev. B {\bf 65}, 140502 (2002).
\bibitem{pedja}
 P. Miranovi\'{c} {\it et al}.,
 Phys. Rev. B {\bf{68}}, 052501 (2003).
\bibitem{izawa}
 K. Izawa {\it et al}.,
 Phys. Rev. Lett. {\bf 89}, 137006 (2002).
\bibitem{tuson}
 T. Park {\it et al}.,
 Phys. Rev. Lett. {\bf 90}, 177001 (2003).
\bibitem{aoki} H. Aoki {\it et al}.,
 J. Phys. Condens. Matter {\bf 16}, L13 (2004).
\bibitem{deguchi}
 K. Deguchi {\it et al}.,
 Phys. Rev. Lett. {\bf 92}, 047002 (2004).
\bibitem{sakata}
 H. Sakata {\it et al}.,
 Phys. Rev. Lett. {\bf 84}, 1583 (2000).
\bibitem{hess}
 H. F. Hess {\it et al}.,
 Phys. Rev. Lett. {\bf 64}, 2711 (1990).
\bibitem{damascelli}
 A. Damascelli {\it et al}.,
 Rev. Mod. Phys. {\bf 75}, 473 (2003).
\bibitem{yokoya} 
 T. Yokoya {\it et al}.,
 Science {\bf 294}, 2518 (2001) .
\bibitem{kiss}
 T. Kiss, thesis (The University of Tokyo, 2004).
\bibitem{ferreira}
 J. Ferreira da Silva {\it et al}., 
 Physica {\bf 41}, 409 (1969).
\bibitem{hanaguri}
 T. Hanaguri {\it et al}.,
 Physica B {\bf 329-333}, 1355 (2003).
\bibitem{boaknin2}
 E. Boaknin {\it et al}.,
 Phys. Rev. Lett. {\bf 90}, 117003 (2003).
\bibitem{hedo}
 M. Hedo {\it et al}.,
 J. Phys. Soc. Jpn. {\bf 67}, 272 (1998).
\bibitem{nohara}
 M. Nohara {\it et al}.,
 J. Phys. Soc. Jpn. {\bf 66}, 1888 (1997).
\bibitem{wang} 
 Y. Wang {\it et al.,}
 Phys. Rev. B {\bf 63}, 094508 (2001).
\bibitem{balci}
 H. Balci and R. L. Greene,
 cond-mat/0402263.
\bibitem{deGennestext}
 See for this concept,  P. G. de Gennes, 
 {\it Superconductivity of Metals and Alloys}
 (W.A. Benjamin, New York, 1966).
\bibitem{volovik}
 G. E. Volovik,
 JETP Lett. {\bf 58}, 469 (1993).
\bibitem{eilenberger}
 G. Eilenberger,
 Z. Phys. {\bf 214}, 195 (1968).
\bibitem{node}
The present ``linear'' line node gives $p\sim 0.5$ 
while the ``quadratic'' node does $p\sim 0.25$. 
Thus we can determine the power behavior of 
the gap function around the node. 
\bibitem{hayashi}
 N. Hayashi {\it et al}.,
 Phys. Rev. Lett. {\bf 77}, 4074 (1996).
\bibitem{pedja2}
 P. Miranovic {\it et al}.,
 cond-mat/0312420.
\bibitem{kita}
 T. Kita,
 cond-mat/0311562.
\bibitem{bouquet}
 F. Bouquet {\it et al}.,
 Phys. Rev. Lett. {\bf 89}, 257001 (2002).
\bibitem{nakai}
 N. Nakai {\it et al}.,
 J. Phys. Soc. Jpn. {\bf 71}, 23 (2002).
\bibitem{ichioka3}
 M. Ichioka {\it et al}.,
 preprint.
\bibitem{nakai2}
 N. Nakai, private communication.
\end{thebibliography}

\end{document}